\documentclass[preprint2]{aastex63}

\usepackage{CJK}
\usepackage{amsmath}
\usepackage{bm}

\newcommand{\petit}{\texttt{petitRADTRANS}}

\newcommand{\pmn}{\texttt{PyMultiNest}}

\newcommand{\caltech}{Department of Astronomy, California Institute of Technology, Pasadena, CA 91125, USA}
\newcommand{\gps}{Division of Geological \& Planetary Sciences, California Institute of Technology, Pasadena, CA 91125, USA}
\newcommand{\ucsc}{Department of Astronomy \& Astrophysics, University of California, Santa Cruz, CA95064, USA}
\newcommand{\keck}{W. M. Keck Observatory, 65-1120 Mamalahoa Hwy, Kamuela, HI, USA}
\newcommand{\ucla}{Department of Physics \& Astronomy, 430 Portola Plaza, University of California, Los Angeles, CA 90095, USA}
\newcommand{\jpl}{Jet Propulsion Laboratory, California Institute of Technology, 4800 Oak Grove Dr.,Pasadena, CA 91109, USA}
\newcommand{\ucsd}{Center for Astrophysics and Space Sciences, University of California, San Diego, La Jolla, CA 92093}

\newcommand{\berkeley}{Department of Astronomy, 501 Campbell Hall, University of California at Berkeley, CA 94720, USA}

\received{\today}
\revised{}
\accepted{}
\submitjournal{ApJ}

\shorttitle{C/O HR 8799 c}
\shortauthors{Wang et al.}
\graphicspath{{./}{figures/}}

\begin{document}
\begin{CJK*}{UTF8}{gbsn}

\title{Retrieving C and O Abundance of HR 8799~\lowercase{c} by Combining High- and Low-Resolution Data}

\correspondingauthor{Ji Wang}
\email{wang.12220@osu.edu}

\author[0000-0002-4361-8885]{Ji Wang (王吉)}
\affiliation{Department of Astronomy, The Ohio State University, 100 W 18th Ave, Columbus, OH 43210 USA}

\author[0000-0003-0774-6502]{Jason J. Wang (王劲飞)}
\altaffiliation{51 Pegasi b Fellow}
\affiliation{Center for Interdisciplinary Exploration and Research in Astrophysics (CIERA) and Department of Physics and Astronomy, Northwestern University, Evanston, IL 60208, USA}.

\author[0000-0003-2233-4821]{Jean-Baptiste Ruffio}
\affiliation{\caltech}

\author{Geoffrey A. Blake}
\affiliation{\gps}

\author{Dimitri Mawet}
\affiliation{\caltech}
\affiliation{\jpl}

\author{Ashley Baker}
\affiliation{\caltech}

\author{Randall Bartos}
\affiliation{\jpl}

\author{Charlotte Z. Bond}
\affiliation{UK Astronomy Technology Centre, Royal Observatory, Edinburgh EH9 3HJ, United Kingdom}

\author{Benjamin Calvin}
\affiliation{\caltech}
\affiliation{\ucla}

\author{Sylvain Cetre}
\affiliation{\keck}

\author[0000-0001-8953-1008]{Jacques-Robert Delorme}
\affiliation{\keck}
\affiliation{\caltech}

\author{Greg Doppmann}
\affiliation{\keck}

\author{Daniel Echeverri}
\affiliation{\caltech}

\author[0000-0002-1392-0768]{Luke Finnerty}
\affiliation{\ucla}

\author[0000-0002-0176-8973]{Michael P. Fitzgerald}
\affiliation{\ucla}

\author[0000-0001-5213-6207]{Nemanja Jovanovic}
\affiliation{\caltech}


\author{Ronald Lopez}
\affiliation{\ucla}

\author[0000-0002-0618-5128]{Emily C. Martin}
\affiliation{\ucsc}


\author{Evan Morris}
\affiliation{\ucsc}


\author{Jacklyn Pezzato}
\affiliation{\caltech}

\author{Sam Ragland}
\affiliation{\keck}

\author[0000-0003-4769-1665]{Garreth Ruane}
\affiliation{\caltech}
\affiliation{\jpl}

\author{Ben Sappey}
\affiliation{\ucsd}

\author{Tobias Schofield}
\affiliation{\caltech}

\author{Andrew Skemer}
\affiliation{\ucsc}

\author{Taylor Venenciano}
\affiliation{Physics and Astronomy Department, Pomona College, 333 N. College Way, Claremont, CA 91711, USA}

\author[0000-0001-5299-6899]{J. Kent Wallace}
\affiliation{\jpl}

\author{Peter Wizinowich}
\affiliation{\keck}

\author{Jerry W. Xuan}
\affiliation{\caltech}

\author{Marta L. Bryan}
\affiliation{\berkeley}

\author[0000-0001-8127-5775]{Arpita Roy}
\affiliation{Space Telescope Science Institute, 3700 San Martin Drive, Baltimore, MD 21218, USA}
\affiliation{Department of Physics and Astronomy, Johns Hopkins University, 3400 N Charles St, Baltimore, MD 21218, USA}

\author{Nicole L. Wallack}
\affiliation{\gps}

\begin{abstract}

The formation and evolution pathway for the directly-imaged multi-planetary system HR 8799 remains mysterious. Accurate constraints on the chemical composition of the planetary atmosphere(s) are key to solving the mystery. We perform a detailed atmospheric retrieval on HR 8799~c to infer the chemical abundances and abundance ratios using a combination of photometric data along with low- and high-resolution spectroscopic data (R$\sim$20-35,000). We specifically retrieve [C/H], [O/H], and C/O and find them to be 0.55$^{+0.36}_{-0.39}$, 0.47$^{+0.31}_{-0.32}$, and 0.67$^{+0.12}_{-0.15}$ at 68\% confidence. The super-stellar C and O abundances, yet a stellar C/O ratio, reveal a potential formation pathway for HR 8799~c. Planet c, and likely the other gas giant planets in the system, {{formed early on (likely within $\sim$1 Myr)}}, followed by further atmospheric enrichment in C and O through the accretion of {{solids beyond the CO iceline. The enrichment either preceded or took place during the early phase of the inward migration to the planet current locations. }}

\end{abstract}



\section{Introduction}
\label{sec:intro}

The HR 8799 multiple planetary system~\citep{Marois2008, Marois2010} remains an interesting and challenging case to test planet formation scenarios. Despite numerous studies, it is still uncertain where and how the four planets formed and how they arrived at their current locations. Chemical composition~\citep[e.g., ][]{Oberg2011, Madhusudhan2017} and orbital configuration~\citep[e.g., ][]{Konopacky2016, Gozdziewski2020} may reveal key information as to the origin and evolutionary history of the system. A comprehensive review can be found in~\citet{Wang2020}. 


{{In~\citet{Wang2020}, we found that retrieval analyses from different groups led to drastically different abundances and C/O ratios. Even with the same retrieval code, the results were sensitive to the choice of priors and data sets. The analyses in~\citet{Wang2020} painted a bleak picture of spectral retrievals for directly-imaged exoplanets. Similar conclusion was also drawn for transiting planets using the James Webb Space Telescopes (JWST), but with an emphasis on the systematics due to the opacity data~\citep{Niraula2022}. However, some of the issues in~\citet{Wang2020} had been addressed with e.g. a more flexible temperature profile and a more realistic cloud treatment~\citep{Wang2022}. Consequently, promising results had been shown when analyzing data from a bench-mark brown dwarfs: consistent chemical abundances and C/O ratios can be retrieved for bench-mark brown dwarfs and their host stars~\citep{Wang2022, Xuan2022}. The promising results motivate a new retrieval analysis for the HR 8799 system.  }}

{{In addition to the progress of spectral analysis, recent observational progress has been made on the study of the HR 8799 planetary system since~\citet{Wang2020}}}.~\citet{Doelman2022} used LMIRCam/ALES observations to further constrain the planetary properties and cloud conditions, {{while~\citet{Sepulveda2022} and~\citet{Brandt2021} further pinned down the dynamical masses for HR 8799 and planets bcde using Gaia data.}}~\citet{Ruffio2021} employed Keck/OSIRIS data to obtain consistent C/O ratios for planets b,c, and d. 

Notably,~\citet{Wang2021} used the Keck Planet Imager and Characterizer~\citep[KPIC, ][]{Mawet2018, Jovanovic2019, Delorme2020,Delorme2021} to obtain high-resolution (R$\sim$35,000) K-band spectra of the four planets orbiting HR 8799. With high-spectral resolution data, they measured previously inaccessible properties such as planet projected rotational velocities. Furthermore, higher spectral resolution can better resolve molecular lines and provide key constraints on chemical compositions even in cloudy conditions~\citep{Xuan2022}.


With the addition of the KPIC data for HR 8799 planets, {{we will again use HR 8799 c to stress test our retrieval analysis and the influence of different data sets on the retrieval results, similar to~\citet{Wang2020, Wang2022}.}} A full analysis of all four planets including the KPIC data and other newly available data will be presented in a forthcoming manuscript. This paper is organized as follows. \S \ref{sec:observation} summarizes the observational data. \S \ref{sec:test} presents a test of our retrieval code using a PHOENIX synthetic spectrum as a known input. Main results are in \S \ref{sec:data_reduction}. Discussions on the results are provided in \S \ref{sec:discussion}. A summary of the paper can be found in \S \ref{sec:summary}.

\section{Data}
\label{sec:observation}

{{Our data for HR 8799 c are detailed in~\citet{Wang2020} and we provide a summary of the data sets in Table \ref{tab:datasets}.}} The data sets include (from lower to higher spectral resolution): $L$- and $M$-band photometric data points~\citep{Bonnefoy2016, Skemer2012, Skemer2014, Currie2014, Galicher2011}, integral field unit (IFU) data from CHARIS (R$\sim$20, PI: Wang), IFU data from GPI~\citep[R$\sim$45-80,][]{Greenbaum2018}, and IFU data from OSIRIS~\citep[R=5000,][]{Konopacky2013}. In addition, we utilize high-resolution spectral data (R$\sim$35,000) for HR 8799 c from KPIC~\citep{Wang2021}. 

\section{Testing with a Phoenix BT-Settl Spectrum}
\label{sec:test}

{{We test our retrieval framework using a synthetic spectrum for which we know the C and O abundance. We refer to~\citet{Wang2022} for details of the retrieval framework because we use the same methodology to perform retrieval analyses. In summary, we model exoplanet atmospheres based on \petit\ and consider both low and high resolution modes (R=1,000 and R=1,000,000). We include MgSiO$_3$ clouds~\citep{Molliere2020} and adopt a flexible P-T profile~\citep{Petit2020}. To sample the posterior distribution in a Bayesian framework, we used \pmn~\citep{Buchner2014}. }}

The synthetic spectrum is obtained from the PHOENIX BT-Settl model~\citep{Baraffe2015}. We choose an effective temperature (T$_{\rm{eff}}$) of 1200 K, a surface gravity ($\log g$) of 3.5, and solar abundances\footnote{The fits file is available at: \url{https://phoenix.ens-lyon.fr/Grids/BT-Settl/CIFIST2011_2015/SPECTRA/}}. The effective temperature and surface gravity of the synthetic spectrum are similar to those of HR 8799 c and the other planets in this system~\citep[e.g., ][]{Doelman2022,Molliere2020}. {{To calculate the luminosity, we set the radius to be 1.20 R$_{\rm{Jupiter}}$. This corresponds to a mass of 1.84 M$_{\rm{Jupiter}}$ for the assumed $\log g$ of 3.5, and a bolometric luminosity of $\log(L/L_\odot)$=-4.62. The luminosity is consistent with the previously measured values of $\log(L/L_\odot)$ from -4.58 to -4.83~\citep{Greenbaum2018} }}

\subsection{Simulating the Data}

{{For low-resolution and photometric data, we resample the synthetic spectrum to the wavelength grid of existing data. We assume a top-hat response function for the photometric filters. To describe the top-hat function for each $L$- and $M$-band photometric filter, we use the central wavelength and the full width at half maximum (FWHM) in Table 1 of~\citet{Bonnefoy2016}. Since the spectral resolution of low-resolution data is not uniform across the wavelength range (as shown in Table \ref{tab:datasets}), sampling the synthetic spectrum to the wavelength grid of existing data ensures that the synthetic spectrum has the same varying spectral resolution as the original spectrum. }}

To simulate high-resolution spectroscopic data, we use a wavelength range from 2.29 to 2.48 $\mu$m. {{We apply a rotational broadening of 10 km$\cdot$s$^{-1}$, which is consistent with the upper limit of 14 km$\cdot$s$^{-1}$ as measured in~\citet{Wang2021}.}} We then convolve the spectrum with a Gaussian kernel that corresponds to spectral resolution of R=35,000, and resample the spectrum to the wavelength grid of the KPIC data. We add random gaussian errors to all simulated data points based on the reported measurement uncertainties.   

We simulate three data sets. The first simulated data set (SDS I) contains only the medium-to-low-resolution and photometric data, {{i.e., data set \#1, \#2, \#3, and \#4 in Table \ref{tab:datasets}. This is the same data set that is used in~\citet{Wang2020}. The second simulated data set (SDS II) contains the high-resolution data (data set \#5). Since the high-resolution data are normalized, we also include simulated CHARIS data (data set \#1) and photometric data (data set \#4) so as to provide flux information for $J$ through $M$ band. The third simulated data set (SDS III) contains simulated data for all observations as listed in \S \ref{sec:observation} and Table \ref{tab:datasets}. The data set includes high- and low-resolution spectral data along with photometric data.    }}

\subsection{Retrieval Analyses For the Simulated Data}

\begin{figure}[h]
\hspace*{-0.4in}
\begin{tabular}{l}
\includegraphics[width=8.5cm]{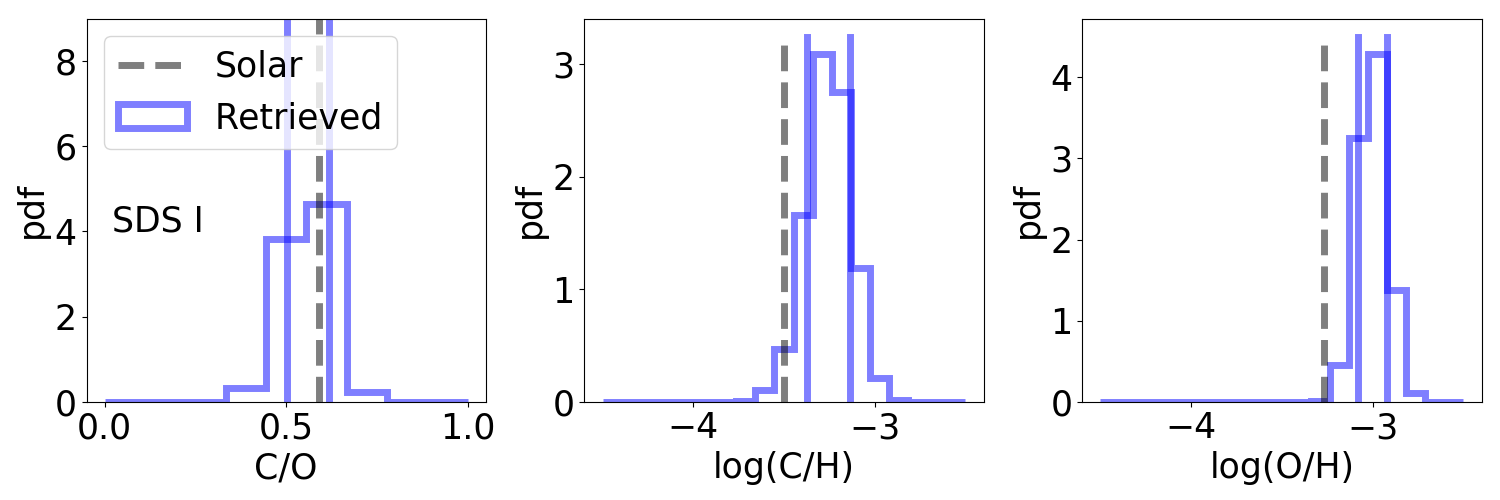} \\
\includegraphics[width=8.5cm]{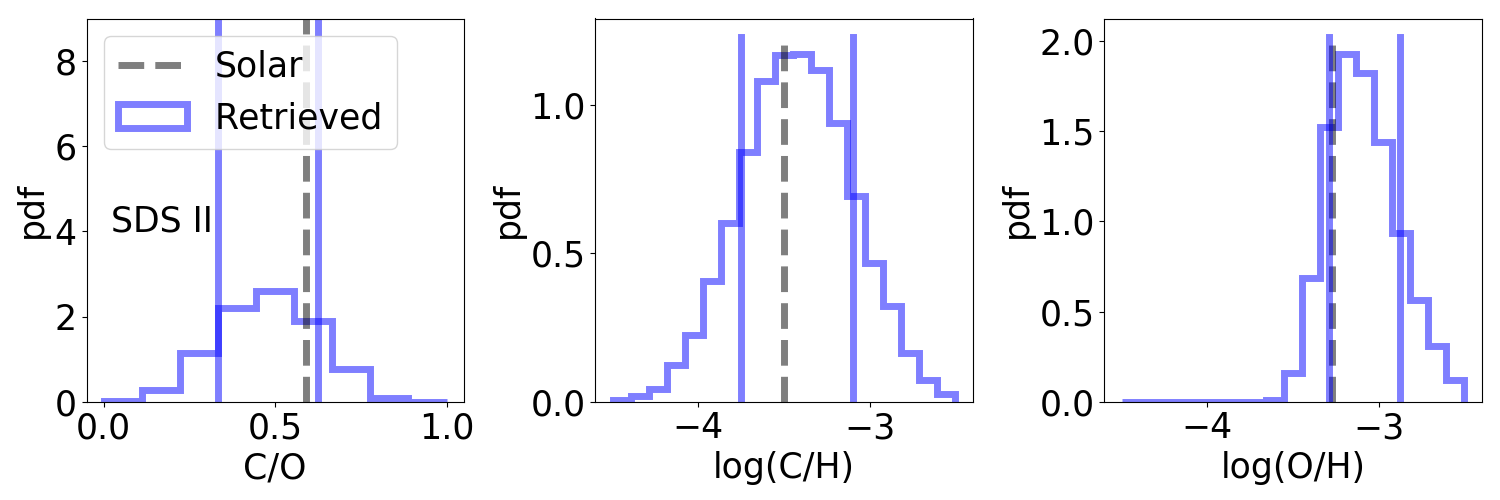} \\
\includegraphics[width=8.5cm]{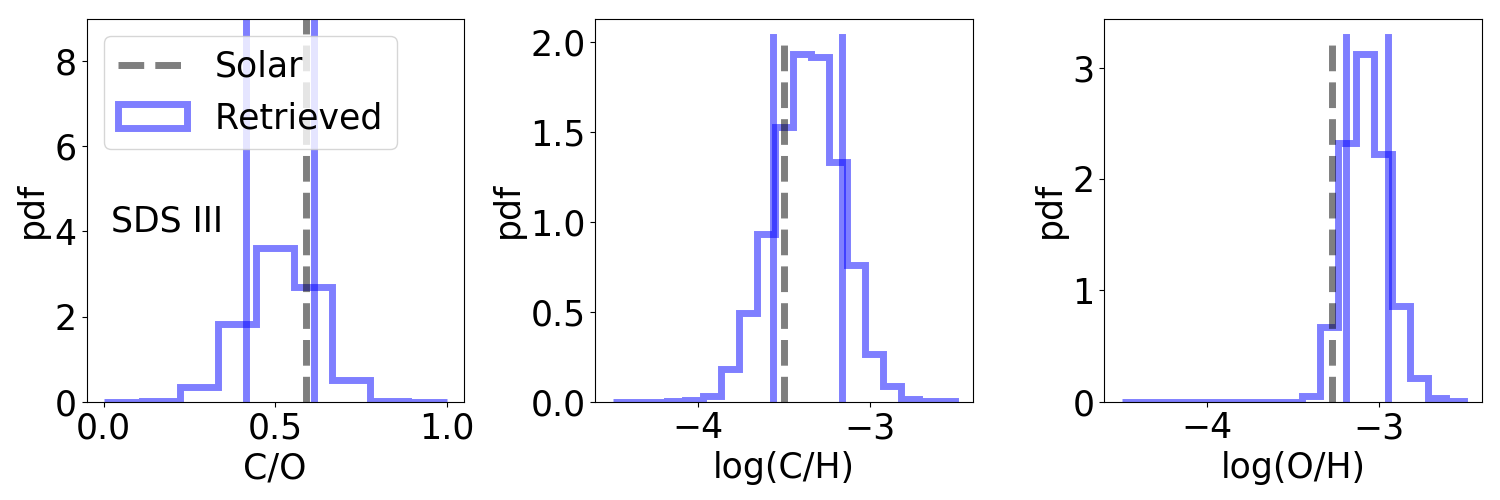} 
\end{tabular}
\caption{{\bf{Retrieved C and O abundances and C/O based on the PHOENIX spectrum.}} {\bf{Top SDS I}}: Low-resolution and photometric data only. {{C and O abundances from posterior samples (blue histograms) tend to be overestimated by 2.6-3.9 $\sigma$ when compared with the solar values~\citep[vertical dashed lines, ][]{Asplund2009}. Vertical blue lines mark 16\% and 84\% percentile values of posterior distributions. }}{\bf{Middle SDS II}}: High-resolution data plus CHARIS and photometric data. C/O is underestimated and C and O abundance uncertainties are the largest. {\bf{Bottom SDS III}}: High and low-resolution along with photometric data. C and O abundances and C/O agree with the solar values within $\sim$1-2 $\sigma$. 
\label{fig:co_comp}}
\end{figure}

The retrieved C and O abundances and C/O ratios are broadly consistent with the solar values, which are the input of the PHOENIX spectrum that we use to simulate the data (Fig. \ref{fig:co_comp}). Retrieved values of model parameters can be found in Table \ref{tab:mcmc_result}. 

{{For SDS I, the [C/H], [O/H], and the C/O ratio are $      0.31^{     +0.12}_{     -0.12}$, $      0.31^{     +0.08}_{     -0.08}$, and $      0.56^{     +0.06}_{     -0.06}$. The C and O abundances are 2.6-$\sigma$ and 3.9-$\sigma$ higher than solar values and the C/O ratio is consistent with the solar value (see Table \ref{tab:CO_abundances} for solar values). 

For SDS II, the [C/H], [O/H], and the C/O ratio are $      0.14^{     +0.33}_{     -0.32}$, $      0.20^{     +0.23}_{     -0.18}$, and $      0.48^{     +0.14}_{     -0.15}$. While the uncertainties are larger, the C and O abundances and the C/O ratio are consistent with the solar values. 

For SDS III, the [C/H], [O/H], and the C/O ratio are $      0.21^{     +0.20}_{     -0.20}$, $      0.24^{     +0.13}_{     -0.12}$, and $      0.52^{     +0.10}_{     -0.10}$. The C and O abundances are 1.0-$\sigma$ and 2.0-$\sigma$ higher than solar values and the C/O ratio is consistent with the solar value within 1-$\sigma$. 

}}

{{Upon further consideration, we decide that combining high- and low-resolution spectral and photometric data offers a balance between accuracy and precision (as demonstrated with SDS III, bottom panel of Fig. \ref{fig:co_comp}). In comparison, retrieval on the data set without high-resolution data (SDS I, top panel of Fig. \ref{fig:co_comp}) significantly overestimates the C and O abundances by 2.6-3.9 $\sigma$. On the other hand, while unbiased, the retrieval analysis on the data set with only high-resolution data and photometric data points (i.e., SDS II) returns the largest uncertainty (middle panel in Fig. \ref{fig:co_comp}). }}

Although the analyses work reasonably well in retrieving C and O abundances, we find that we cannot satisfactorily retrieve the input $\log g$. {{The input value for $\log g$ is 3.5, but we consistently retrieve a $\log g$ that is higher than 4.2, which is $>$3-$\sigma$ higher than the input value. In comparison, the retrieved planet radius is generally overestimated by $\sim$2-$\sigma$ when compared to the input value of 1.2 R$_{\rm{Jupiter}}$. The retrieved T$_{\rm{eff}}$ is consistent with the input value (1200 K) within $\sim$1-$\sigma$: the retrieved values 
of T$_{\rm{eff}}$ from SDS I, II, and III are $      1228^{     +126}_{     -40}$ K, $      1215^{     +69}_{     -42}$ K, and $      1300^{     +253}_{     -95}$ K, respectively. }}While the retrieved $\log g$ is inconsistent at $>$3-$\sigma$ level, this does not affect the main conclusion of the paper, which rests on the C and O abundances and the C/O ratio of HR 8799~c (see \S \ref{sec:fixingLoggRadius} for more discussion).  




\begin{figure*}[h]
\begin{tabular}{c}
\includegraphics[width=14.0cm]{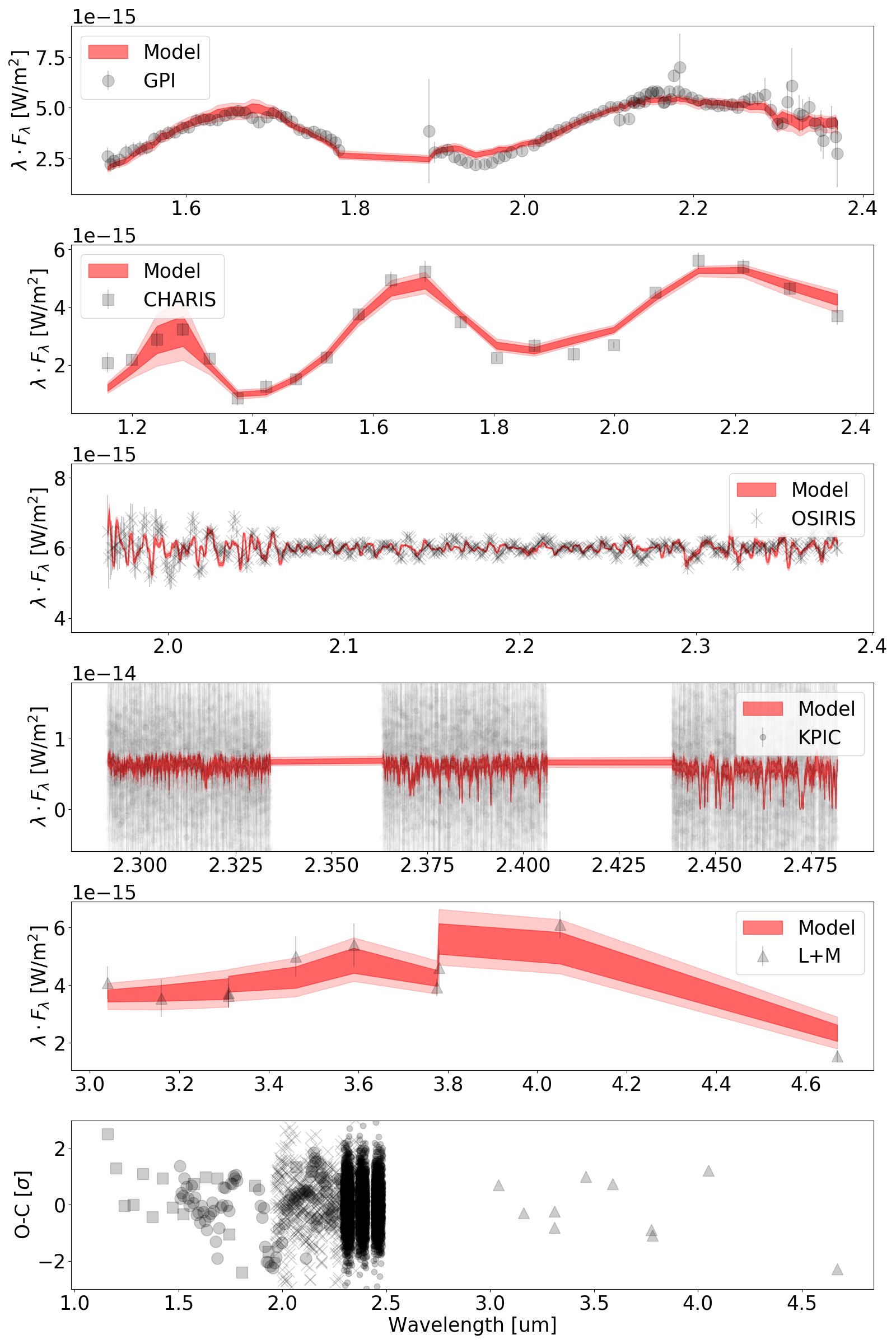}
\end{tabular}
\caption{{\bf{Retrieved results for HR 8799 c}}. Top five panels are the observed spectroscopic and photometric data (black) and the 1-$\sigma$ (16 to 84 percentile, darker red) and 2-$\sigma$ (2.5 to 97.5 percentile, lighter red) distribution of modeled spectra. OSIRIS and KPIC spectral data are normalized so the flux on the y-axis is arbitrary. The bottom panel is a residual plot with data minus model and divided by the individual errors. 
\label{fig:data_model_hr}}
\end{figure*} 

\begin{figure*}[h!]
\epsscale{1.0}
\plotone{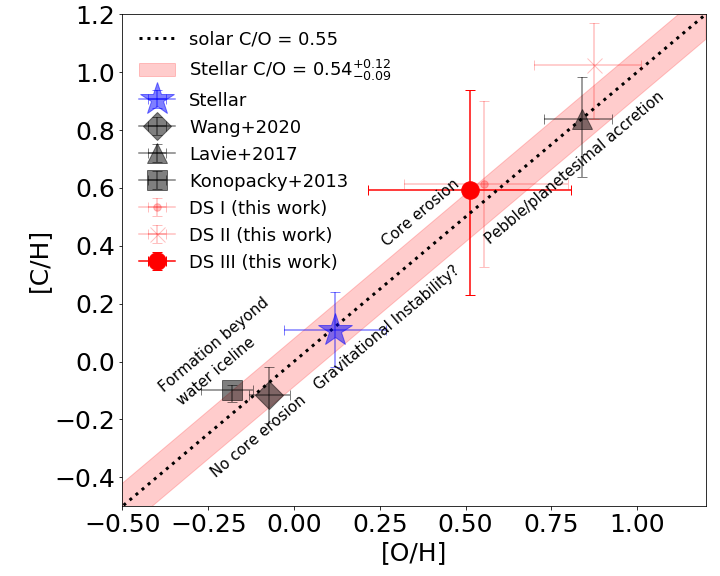}
\caption{Carbon and oxygen abundance for HR 8799 (blue) and planet c (black and red data points). The dashed line represents solar C/O~\citep[0.55, ][]{Asplund2009} and the red shaded region is the 1-$\sigma$ uncertainty region for HR 8799 stellar C/O (0.54$^{+0.12}_{-0.09}$). The quoted values for HR 8799 c are obtained from the retrieval analysis on DS III which include low- and high-resolution spectral data as well as photometric data. See \S \ref{sec:low_high_comp} for more details for DS III and other comparison data sets (DS I and DS II).
\label{fig:co}}
\end{figure*}

\section{Key Results From the Retrieval Analysis for HR 8799 \lowercase{c}}
\label{sec:data_reduction}

We focus on the retrieval analysis on all available data for HR 8799 c (DS III in Table \ref{tab:mcmc_result}) because our retrieval results on simulated data sets in \S \ref{sec:test} shows that combining high- and low-res spectral and photometric data points leads to an accurate and precise measurement of C and O abundances.

Fig. \ref{fig:data_model_hr} shows that our model fits reasonably well for all available data: our retrieved model spectra fit data within 3-$\sigma$ (bottom panel in Fig. \ref{fig:data_model_hr}). Noticeable mismatches include the short-wavelength end of the $J$-band CHARIS data, shorter wavelength to 2 $\mu$m in the $K$-band GPI data, and the $M$-band photometric data. These mismatches were also seen in~\citet{Wang2020}.   

We retrieve C and O abundances and the C/O ratio that are higher than (at $\sim$1-$\sigma$ level) the host star values (Table \ref{tab:CO_abundances}). For HR 8799 c, the [C/H], [O/H], and C/O are 0.55$^{+0.36}_{-0.39}$, 0.47$^{+0.31}_{-0.32}$, and 0.67$^{+0.12}_{-0.15}$ at 68\% credible interval. The 68\% credible interval applies to all quoted uncertainties in this paper. In comparison, the host star HR 8799 [C/H], [O/H], and C/O are 0.04$\pm$0.12, 0.08$\pm$0.14, and 0.54$^{+0.12}_{-0.09}$~\citep{Wang2020}. 

{{It is important to note that our test on SDS III in \S \ref{sec:test} suggests that the retrieved [C/H] and [O/H] may be overestimated by 0.21 and 0.24 dex and the C/O ratio underestimated by 0.03. Assuming the same bias exist for the analysis on DS III and after correcting for the bias, the [C/H], [O/H], and C/O are 0.34$^{+0.36}_{-0.39}$, 0.23$^{+0.31}_{-0.32}$, and 0.70$^{+0.12}_{-0.15}$ for HR 8799 c. As a result, the conclusion of super-stellar abundance is weakened and super-stellar C/O is strengthened. However, the super-stellar abundances and C/O ratio remain at $\sim$1-$\sigma$ level. }}

\subsection{Formation Pathway}
\label{sec:formation}

\subsubsection{A Brief Introduction}

Fig. \ref{fig:co} shows the 2-D parameter space of [C/H] vs. [O/H], which may be informative as to the planet formation pathway~\citep{Madhusudhan2017}. For example, if the planet and the host star share similar chemical abundances, then it is likely that they form in a similar fashion to a binary star via gravitational instability that leads to fragmentation in a molecular cloud. Alternatively, a significant difference in chemical abundances and elemental ratios would indicate formation scenarios that are more akin to core accretion pathway. 

As summarized in~\citet{Oberg2011}, when migration is minimal a stellar C/O ratio is expected for planets forming interior to both the water iceline and the carbon-grain evaporation line, where all volatiles are in the gas phase. Sub-stellar C/O and sub-stellar abundance indicate formation beyond the water iceline but inside the CO iceline. Substellar or stellar C/O and super-stellar absolute abundances (relative to hydrogen) indicate that a large amount of solids ``polluted” the atmosphere following gas accretion. Super-stellar C/O and abundances may be due either to gas accretion close to the CO or CO$_2$ icelines, or indicate that the atmosphere was substantially polluted by pebble (early on) or planetesimal (above the pebble isolation mass) accretion and/or core erosion. When migration is significant, the various environments along the path can contribute, leading to less stringent constraints on planet formation~\citep{Pacetti2022}.

Results from~\citet{Konopacky2013} and~\citet{Wang2020} are consistent with a scenario in which the planet forms between the H$_2$O and CO$_2$ icelines, accreting gas that is deficient in H$_2$O, and core erosion plays no role in enriching the planet atmosphere. In comparison, the result from~\citet{Lavie2017} points to a scenario in which the planet atmosphere is significant affected by pebble/planetesimal accretion and/or core erosion post formation. 

\subsubsection{One Likely Scenario}

Our new result---super-stellar/stellar C/O and super-stellar metallicity---suggests the following formation scenario: the HR 8799 Jovian planets formed {{well before $\sim$1 Myr}}, with their atmospheres contaminated by C and O through accretion of {{solids}}, and the planets migrated inward to their current locations. 

First, we address the formation {{time scale. The formation of the planets need to be very early to be consistent with the observed level of metal contamination. It is estimated that 65-360 M$_\oplus$ of solids have been accreted to account for the observed $\sim$0.5 dex metal enrichment~\citep{Molliere2020}. The large solid budget is only possible to be available during the Class I phase~\citep{Najita2014}, which is much younger than 1 Myr. The early planet formation is also supported by evidence of planet formation in embedded phases for young stars~\citep{Harsono2018}. 

Forming multiple planets early on can be challenging for both gravitational instability and core accretion. While gravitational instability can form companions on a time scale that is less than $\sim$0.1 Myr~\citep{Pineda2015}, there is little evidence that the instability can form planets in orbital resonance, which is a sign of orbital migration for HR 8799 planets~\citep{Konopacky2016,Wang2018}. Early work also disfavors the core-accretion scenario given the insufficient mass budget and short formation time scale~\citep{DodsonRobinson2009}. However, recent studies show that gas giant planets can form within $\sim$1 Myr via streaming instability~\citep{Youdin2005,Li2019} and pebble accretion~\citep{Ormel2010,Johansen2017}.  }}

Second, we address the accretion of {{solids. The accreted solids can be in the form of dust, pebbles, and/or planetesimals. In the core accretion scenario via streaming instability and pebble accretion, the accreted solids can be dominated by pebbles. The key difference between pebbles and other smaller dust or larger planetesimals is that pebbles feel the gas drag and move toward locations of pressure maxima. In the case of forming planets via gravitational instability, as planets form early and open gaps, pebbles can no longer be accreted onto the growing planets. In comparison, accretion of planetesimals and small dust can ensue through diffusion and dynamical scattering. }}


Lastly, we address the inward orbital migration and atmospheric contamination. In accretion models, only solid accretion outside the CO iceline, where solids have stellar C/O, would result in a stellar C/O in the planet atmosphere. At other radial locations, the planet C/O can vary greatly. For example, the C/O ratio be significantly lower than stellar C/O, e.g., accretion of solids that are enriched by O but not C (e.g., H$_2$O) within the CO iceline but outside the water ice line. ~\citet{Molliere2020} shows that the location for CO iceline ($\sim$45 AU) is much further away than the current location of HR 8799~e ($\sim$15 AU). Therefore, under this scenario, HR 8799~e would have to migrate inward and other planets (including planet c) would have to migrate inward too in order to maintain the resonant chain. 

Moreover, the C and O abundances and C/O for HR 8799~e~\citep{Molliere2020} and 8799~c in this work have comparable C and O abundances and C/O. Both planets have super-solar metallicities at $\sim$0.5 dex and a C/O that is consistent with the stellar value. For 8799~e, it is estimated that 65-360 M$_\oplus$ of solids have been accreted~\citep{Molliere2020}. The amount of accreted solids for 8799~c should be comparable. 

The timing of the atmospheric pollution should precede the inward migration or take place during the early phase of the migration when the planets are outside the CO iceline. This is to ensure that the majority of the accreted solids have stellar C/O ratio. However, accretion of gas and solids from the infalling protostellar envelope may complicate the interpretation. Depending on the original formation location, the gas/dust ratio for the infalling material can vary significantly~\citep[e.g.,][]{Visser2009}. Thus, detailed modeling works of the HR 8799 exoplanetary atmospheric compositions that involve infalling gas+dust, accretion, and migration should provide further constraints on their formation and migration history.  


\subsubsection{Unlikely Scenarios}

In principle, gas accretion close to the CO iceline can also lead to stellar/super-stellar CO and super-stellar metallicity~\citep{Oberg2011}, as the evaporation front close to the CO iceline results in an overabundance of CO. Therefore, the accretion of CO gas would lead to stellar C/O and super-stellar abundances that we measured. From~\citet{Oberg2011}, CO/H$_2$ gas enrichment by a factor of two would lead to a stellar C/O and an enrichment of 2 times (0.3 dex). A higher level of CO enrichment would result in a higher level of metal-enrichment. However, in multiple-planet systems such as the HR 8799 system, it is unlikely that both planet c and e form in the same region where CO abundance is enhanced. 

In a different scenario in which planets form inside the water ice line but outside the refractory carbon stability front, the C/O ratio can approach unity from the accreted solids (Chan et al., submitted). This would require outward migration for the 8799 planets. However, we consider this scenario unlikely given the insufficient mass budget in forming multiple planets between the iceline and the soot line.   

We also discuss a core-erosion scenario~\citep{Madhusudhan2017}, but conclude that the scenario is unlikely given the high-level of metal-enrichment of the atmosphere. 

In the core-erosion scenario, the post-formation atmospheric contamination is achieved by core erosion. The level of enrichment depends on how much metal can be provided by the core. In the core-accretion planet formation scenario, the mass threshold for the core is $\sim$10 M$_\oplus$, so the core can provide up to $\sim$10 M$_\oplus$ metals. This is insufficient when compared to the amount of solids~\citep[65-360 M$_\oplus$, ][]{Molliere2020} that is needed to enrich the atmosphere.

In the disk-fragmentation scenario, the core mass is estimated to be $\sim$15 M$_\oplus$ assuming a planet mass of 5 M$_{\rm{Jupiter}}$ and a dust-to-gas ratio of 1 to 100. While the dust-to-gas ratio can be enhanced in a proto-planetary disk, the core mass seems insufficient to enrich the atmosphere at the observed level.




\subsection{Temperature Structure and Cloudiness}
\label{sec:pt}

\begin{figure*}[ht]
\epsscale{1.1}
\plotone{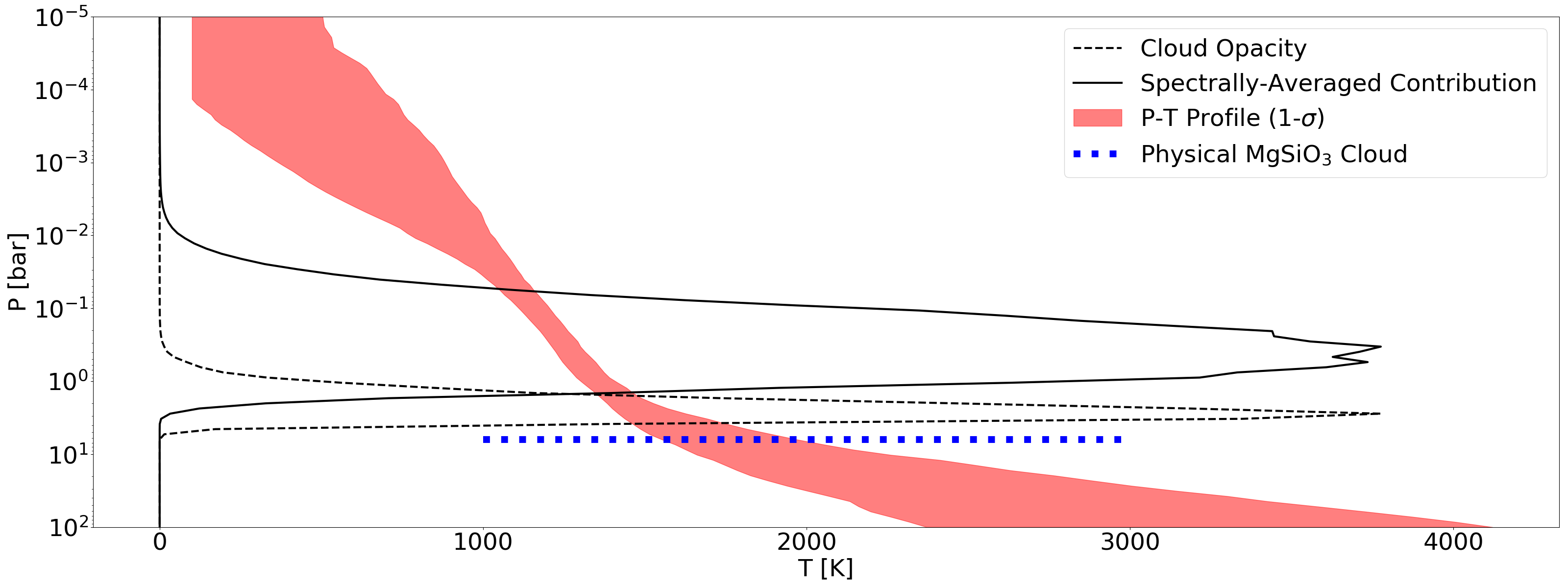}
\caption{{\bf{Retrieved P-T profile (1-$\sigma$ region in red shaded region) for HR 8799 c using DS III.}} The data set (DS III) include low- and high-resolution spectral data as well as photometric data. See \S \ref{sec:low_high_comp} for more details for this data set and other data sets (DS I and DS II). The spectrally-averaged contribution function is shown as the black solid line, which partially overlaps with retrieved cloud layer (black dashed line). The pressure level of the retrieved cloud layer is consistent with that of a physical MgSiO$_3$ cloud (blue dotted line). 
\label{fig:pt}}
\end{figure*}

The retrieved P-T profile is shown in Fig. \ref{fig:pt}. The contribution function (black solid line in Fig. \ref{fig:pt}) shows that the majority of the thermal flux comes from pressures between 0.1 and 1 bar, at a temperature of $\sim$1200 K. This is consistent with the effective temperature of the planet~\citep{Wang2020}. Below $\sim$10 bar, the P-T profile has a thermal gradient ($d\log T/d\log P$) about 0.25, indicating a convective layer. Between 0.1 and 10 bar, the thermal gradient decreases, implying that the dominant energy transporting mechanism shifts from convection to radiation. 

At lower pressure levels (P$<$0.1 bar), the P-T profile is less constrained, but the increased thermal gradient may indicate some detached convective regions as suggested in~\citet{Marley2015}. The upper convective regions can facilitate vertical mixing that has been suggested to cause chemical disequilibrium for many brown dwarfs and exoplanets, including HR 8799 c~\citep[e.g., ][]{Skemer2014}.

The contribution function partially overlaps with the cloud opacity (dashed line in Fig. \ref{fig:pt}). {{Below we provide a calculation of cloud optical depth and discuss the implication of the cloudiness of HR 8799 c. The retrieved cloud opacity is $\sim$10$^{-4}$ cm$^2$/g with a long tail towards to lower values. At the cloud base, the pressure is $\sim$10 bar and the temperature is $\sim$1600 K. The density $\rho$ is 2$\times10^{-4}$ g/cm$^3$ assuming an ideal gas law with a mean molecular weight of 2.3. The cloud spans $\sim$1.5 dex of pressure change (from 10 bar to 0.3 bar as shown in Fig. \ref{fig:pt}. This corresponds to $\sim$6 scale heights. At a temperature of $\sim$1600 K and a $\log g$ of 4.2 with a mean molecular weight of 2.3, each scale height is 36 km. Therefore, 6 scale heights are 216 km. Based on $\tau=\kappa\rho l$, where $\tau$ is optical depth, $\kappa$ is opacity, and $l$ is the cloud thickness, the optical depth is 0.4.  }}

{{The optical depth is comparable with unity. Note that the retrieval model is one-dimensional and retrieves the average of a 3D object. This means that certain regions would have cloud optical depth higher than 1 given the patchiness of clouds. Therefore, it is likely that clouds play a role in affecting the emerging thermal flux even though the cloud contribution function is slightly below the flux contribution function. Our inference of the cloudiness is consistent with previous observational and modeling work for HR 8799 planets~\citep[e.g., ][]{Skemer2014,Molliere2020, Skemer2012, Currie2014}. }}

\section{Discussion}
\label{sec:discussion}

\subsection{Comparing C and O Abundances From Different Data Sets}
\label{sec:low_high_comp}

{{In addition to the retrieval analysis using all available data for HR 8799 c, which include high- and low-spectral resolution along with photometric data points (i.e., DS III, similar to the nomenclature in \S \ref{sec:test}), we also run separate analyses using different combination of data sets. }}The results are shown in Table \ref{tab:mcmc_result}. The choices to construct a data set for retrieval are similar to those in \S \ref{sec:test}. One data set (DS I hereafter) uses all data but the high spectral resolution data from KPIC. The other data set (DS II hereafter) uses high spectral resolution data from KPIC, CHARIS data that covers $J$ through $M$-band, and photometric data points.  

\begin{figure}[h]
\hspace*{-0.4in}
\begin{tabular}{l}
\includegraphics[width=8.5cm]{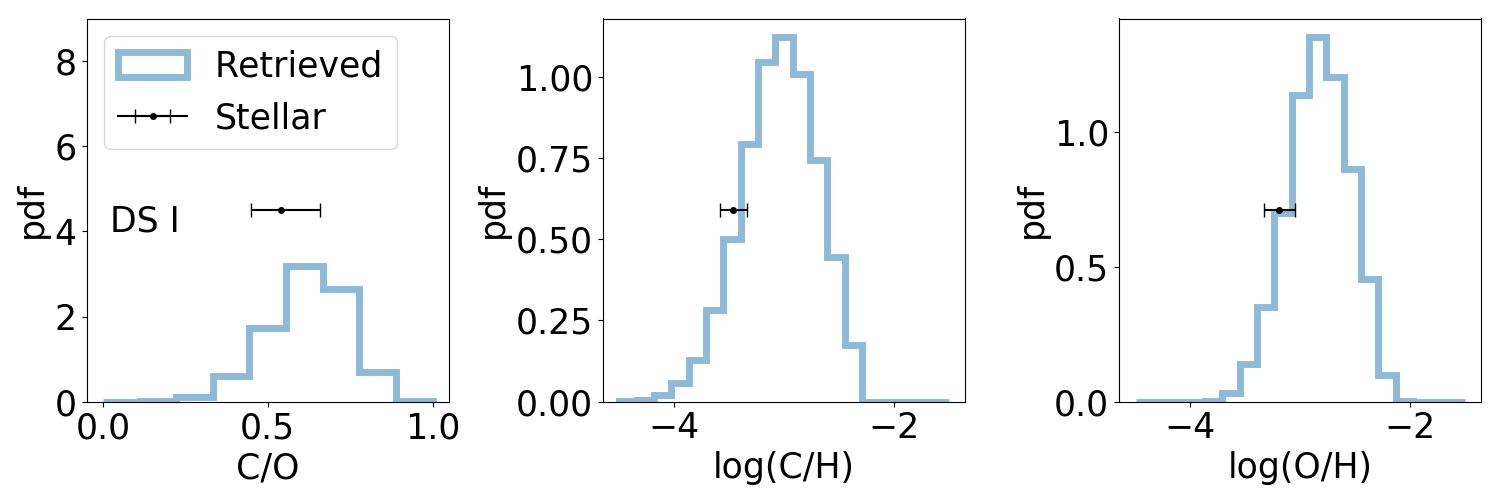} \\
\includegraphics[width=8.5cm]{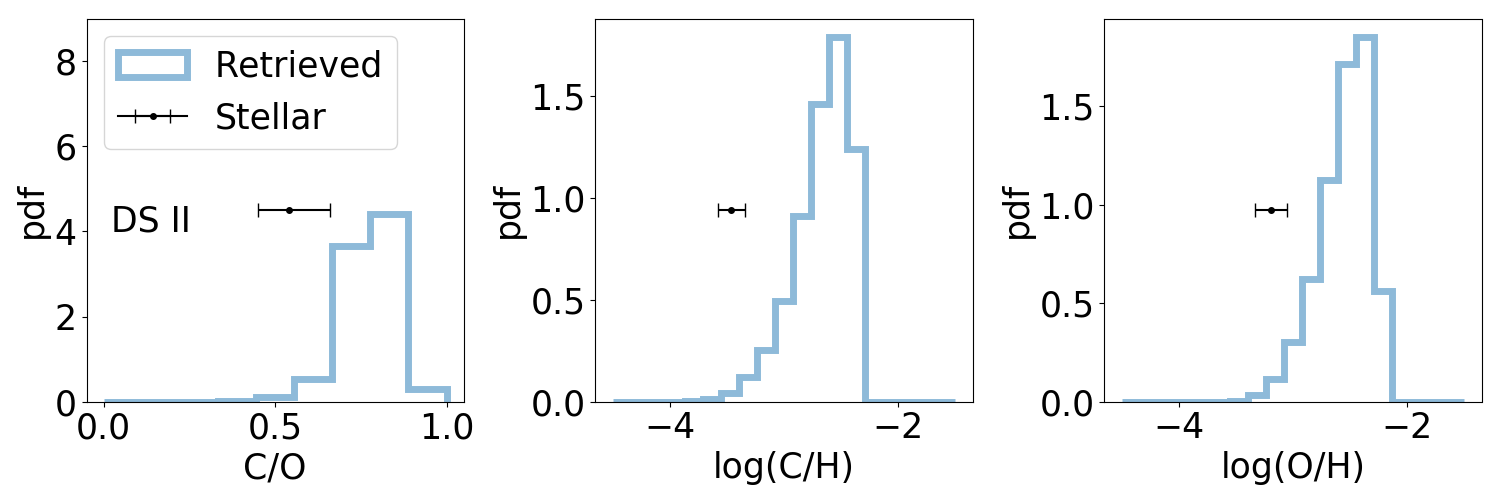} \\
\includegraphics[width=8.5cm]{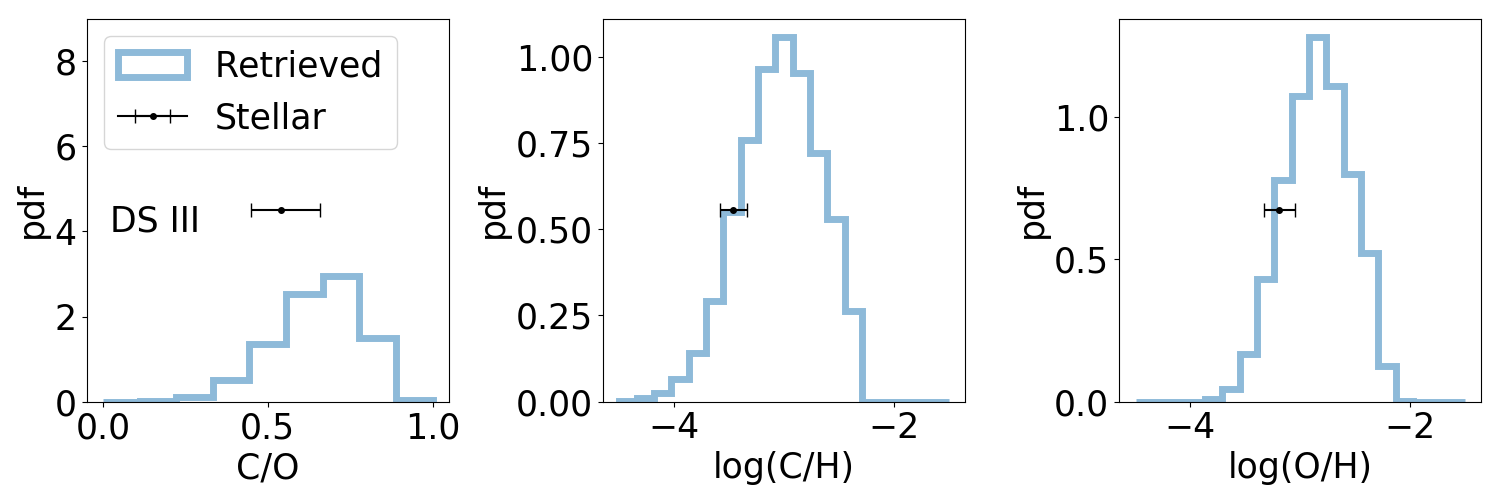} 
\end{tabular}
\caption{{\bf{Retrieved C and O abundances and C/O based on the HR 8799 c data: C and O abundances from posterior samples are blue histograms, and stellar values are the black data points.}} {\bf{Top}}: DS I: Low-resolution and photometric data only. {\bf{Middle}}: DS II: High-resolution data plus CHARIS and photometric data. {\bf{Bottom}}: DS III: High and low-resolution along with photometric data. The difference between DS and SDS in Fig. \ref{fig:co_comp} is that DS indicates real data and SDS indicates simulated data.
\label{fig:co_comp_hr_comp}}  
\end{figure}

The most interesting case is DS II. The analysis on DS II returns much higher C and O abundances and C/O than that for DS I and DS III (Fig. \ref{fig:co_comp_hr_comp}). In addition, the retrieved uncertainties are the smallest among all DS data sets. This is not seen in our analysis on the simulated data set SDS II. 

Moreover, DS II is the data set with the most constrained f$_{\rm{blur}}$ term at $     12.08^{     +4.23}_{     -2.89}$ km$\cdot$s$^{-1}$. The term consists of both the spectral blurring due to a finite resolving power and the rotational broadening. {{Assuming the spectral blurring term and rotational broadening term sum up in quadrature to equal to the measured f$_{\rm{blur}}$ term, if the spectral blurring is $\sim$8.5 km$\cdot$s$^{-1}$ at R=35,000, the rotational broadening is $\sim$8.7 km$\cdot$s$^{-1}$ with a sizable uncertainty of 3.0-4.5 km$\cdot$s$^{-1}$. }}This is consistent with the upper limit of 14 km$\cdot$s$^{-1}$ given in~\citet{Wang2021}. DS III constrains f$_{\rm{blur}}$ to be $\sim$20 km$\cdot$s$^{-1}$, which is also consistent with the the upper limit in~\citet{Wang2021} given the large uncertainty (Table \ref{tab:mcmc_result}). {{DS I does not constrain f$_{\rm{blur}}$ because the low-resolution data have no constraining power on f$_{\rm{blur}}$: OSIRIS data with the highest spectral resolution (R=5000) among low-resolution data can only resolve f$_{\rm{blur}}$ as low as $\sim$60 km$\cdot$s$^{-1}$, which is not sufficient for the $\sim$10 km$\cdot$s$^{-1}$ level for HR 8799 c. }}

Overall, we find that DS III and DS I return similar C and O abundances and C/O. This is not surprising given that the most constraining power comes from the lower-resolution data sets. We follow~\citet{Xuan2022} to quantify the relative constraining power between the high-resolution KPIC and the lower-resolution IFU data. {{The detection significance using KPIC high-resolution data is $\sim$10~\citep{Wang2021}. This detection significance is quantified by the significance of the peak of a cross correlation function (CCF) when cross-correlating the KPIC spectrum with a template. The detection significance using CCF is also commonly referred to as CCF signal-to-noise-ratio (SNR). In comparison, median SNR per wavelength bin is typically higher than 10 for lower-resolution date sets (Table \ref{tab:datasets}). }}Therefore, the lower-resolution data sets dominate the constraining power, DS I and DS III return similar results because lower-resolution date sets are used in these two data sets. 

Based on our tests on the synthetic PHOENIX spectrum, we adopt the retrieval result using DS III. Our key findings in \S \ref{sec:data_reduction} remain unchanged for DS I-III even with the high C and O abundances and C/O retrieved for DS II (Fig. \ref{fig:co}).

\subsection{Comparing Retrieved Uncertainties}
\label{sec:unc}

Properly estimated uncertainties enable accurate interpretation of the retrieval results. Below we compare uncertainties in our retrieval analyses.  

The uncertainties for the HR 8799 c data analyses (DS I and III) are generally $\sim$2-3 time higher than those for the PHOENIX analyses (SDS I and III). We attribute the smaller uncertainties for SDS I and III to the similarities of underlying physics of radiative transfer in atmospheres between \petit\ and PHOENIX. The larger uncertainties in DS I and III can be interpreted as accumulative differences between modeling and real atmospheres as marginalized over many modeling parameters. 

DS I was used for retrieval analysis in~\citet{Wang2020}, and is re-analyzed using our updated retrieval code~\citep{Wang2022}. The uncertainties for [C/H], [O/H], and C/O are 0.1 dex, 0.06 dex, and 0.05 in ~\citet{Wang2020}. In our new analysis on DS I, the uncertainties for [C/H], [O/H], and C/O are $\sim$0.35 dex, 0.3 dex, and 0.1. The new uncertainties are 2-5 times larger than those from the previous analysis. We attribute the increased uncertainties, and likely more realistic, to the updated and more flexible model, e.g., cloud treatment and P-T profile, as detailed in \S 4.2 in~\citet{Wang2022}. 

DS II represents an interesting case as discussed in \S \ref{sec:low_high_comp}. A relevant case is the analysis on HR 7672 B~\citep{Wang2022} where the data set is similarly a combination of high-resolution spectral data and photometric data points. The uncertainties for [C/H] and [O/H] are $\sim$0.05 dex and C/O uncertainty is 0.02. These values are much lower than those for DS II in this work ($\sim$0.2 dex for C and O abundances and 0.06 for C/O). The higher uncertainties for DS II are due to the much lower SNR (SNR$\sim$1 per pixel) of the HR 8799~c high-resolution data than the HR 7672 B data (SNR$\sim$10-20 per pixel). 

DS II analysis also returns the lowest uncertainties among DS I-III. One possible explanation is the more homogeneous data used in DS II. More data from inhomogeneous sources can introduce difference systematics and therefore broadens the posterior distribution~\citep{Wang2020}. 

An alternative explanation is that high spectral resolution data are more effective in constraining chemical abundances and atmospheric properties~\citep[for HD 4747 B,][]{Xuan2022}. However, in the case of HR 8799~c, we prefer results using DS III over DS II for the following reasons. First, our test using PHOENIX spectrum shows that combining high- and low-resolution spectral data as well as photometric data points leads to the most accurate results (\S \ref{sec:test}). Second, our low-resolution spectral and photometric data covers a large wavelength from $J$ to $M$-band. The large wavelength coverage includes spectral range from 2.2 to 2.5 $\mu$m and the $L$ and $M$-band, which are thought to be crucial in constraining chemical abundances~\citep{Xuan2022} and cloud properties~\citep{Wang2020}. This is a key difference between the HD 4747 B data and the HR 8799~c data. Lastly, the major conclusions of the paper would remain the same even if we adopt the retrieval result from DS II.  


\subsection{The Influence of Radius and Gravity on C and O Abundances}
\label{sec:fixingLoggRadius}

As noted in \S \ref{sec:test} for the synthetic data, the retrieved radius ($\sim$1.3 R$_{\rm{Jupiter}}$) and surface gravity ($\log g\sim$4.2) are off by more than 3-$\sigma$ from the input values of the synthetic PHOENIX spectrum, which are 1.2 R$_{\rm{Jupiter}}$ for radius and 3.5 for $\log g$. 

We run a separate retrieval analysis to investigate the influence of radius and surface gravity on C and O abundances (see results in Table \ref{tab:mcmc_result}). We compare the results from the analysis in \S \ref{sec:test} to those for which we fix the radius and $\log g$ at 1.2 R$_{\rm{Jupiter}}$ and 3.5. We find that the retrieved C and O abundances and C/O are consistent between these retrieval analyses. Changing the radius and surface gravity from 1.2 and 3.5 to 1.3 and 4.2 shifts the C and O abundance upward by $\sim$0.2 dex and the C/O downward by 0.03. 

These changes are smaller than, but not negligible when compared to, the uncertainties in the C and O abundances and the C/O ratio. If accounting for the effect of surface gravity on C and O abundances, [C/H] and [O/H] for HR 8799~c would shift downward by $\sim$0.2 dex if $\log g$ drops from 4.2 to 3.5, but the major conclusions in \S \ref{sec:data_reduction} remain unchanged. We also note that the retrieved radius at $1.01^{+0.09}_{-0.08}$ R$_{\rm{Jupiter}}$ and $\log g$ at $4.17^{+0.41}_{-0.47}$ for HR 8799~c (DS III in Table \ref{tab:mcmc_result}) correspond the a planetary mass of $\sim$6 M$_{\rm{Jupiter}}$. These values agree well with most recent literature values~\citep{Doelman2022} and the mass is unlikely cause the long-term dynamical instability of the system~\citep{Wang2018b} . 

\subsection{Chemical Disequilibrium}
\label{sec:diseq}

The atmosphere of HR 8799 c is known to be in chemical disequilibrium~\citep[e.g., ][]{Skemer2012,Konopacky2013}. Here, we investigate (1) if our results are consistent with the chemical disequilibrium case; and (2) to what extent we can constrain the vertical mixing which is thought to disrupt equilibrium chemistry. 

We use \texttt{poor\_mans\_nonequ\_chem} to interpolate a pre-calculated chemical grid from \texttt{easyCHEM}~\citep{Molliere2017}. The grid spans multiple dimensions including temperature (60 - 4000 K), pressure (10$^{-8}$ - 1000 bar), C/O (0.1 - 1.6) and [Fe/H] (-2 - 3). To calculate the equilibrium abundance, we use the median of the retrieved P-T profile and stellar values of C/O = 0.67 and [Fe/H] = 0.50. The value for the metallicity [Fe/H] is chosen to be consistent with the retrieved [C/H] and [O/H] at 0.55 dex and 0.47 dex, respectively (see Table \ref{tab:mcmc_result} for DS III results). 

\begin{figure*}[h!]
\epsscale{1.1}
\plotone{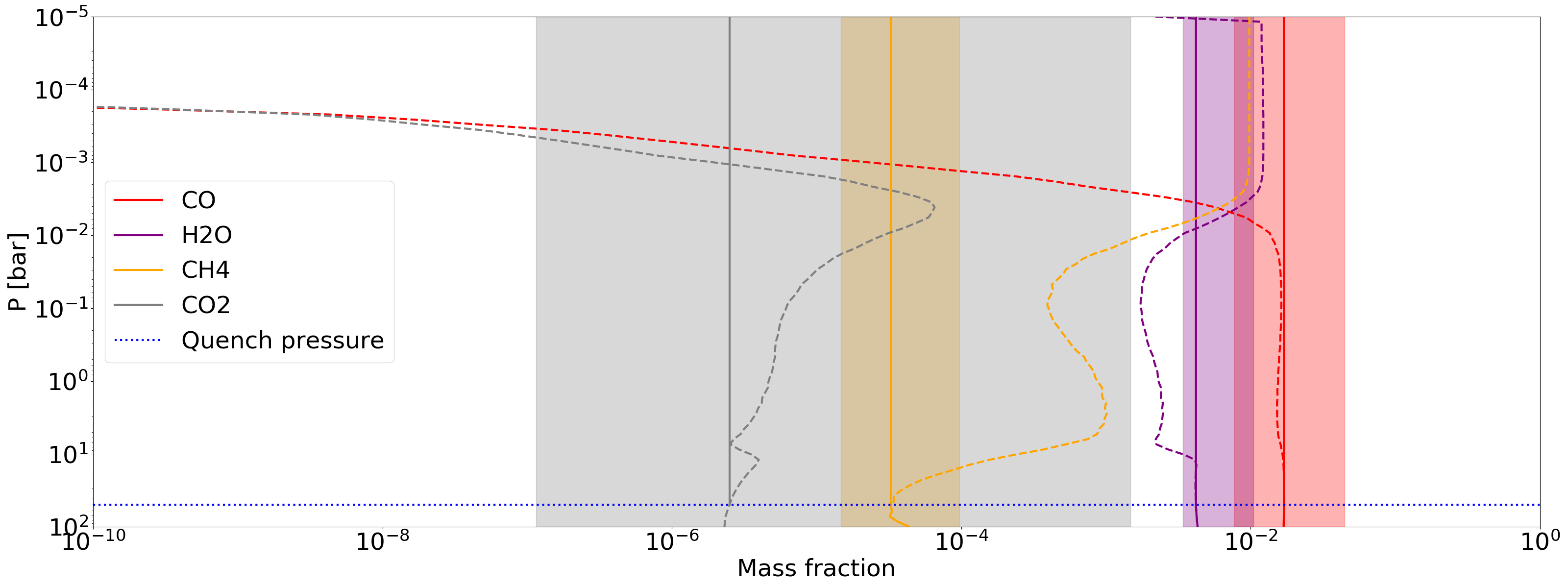}
\caption{Abundances assuming equilibrium chemistry (dashed lines) and a quenched pressure at 50 bar (the horizontal blue dashed line). 1-$\sigma$ ranges of retrieved abundances are shown in shaded regions. The retrieved abundances are inconsistent with the predicted values assuming equilibrium chemistry, at pressures between 0.1 and 1 bar which is the peak of the contribution function (see Fig. \ref{fig:pt}). Instead, the retrieved abundances are consistent with the predicted values assuming a perfect vertical mixing for pressures lower than 50 bar (vertical solid lines). This indicates a case of chemical disequilibrium and a quench pressure that is higher than $\sim$10 bar. 
\label{fig:chem_equi}}
\end{figure*}

Fig. \ref{fig:chem_equi} shows the abundances (in mass mixing ratio) for four species (CO, H$_2$O, CH$_4$, and CO$_2$) assuming two conditions: chemical equilibrium, and a quenched case of chemical disequilibrium, where vertical mixing homogenizes abundances above a quench pressure which we set at 50 bar.

The retrieved abundances are inconsistent with the predicted values assuming equilibrium chemistry. At pressures between 0.1 and 1 bar, which is the peak of the contribution function (see Fig. \ref{fig:pt}), the abundances as predicted from equilibrium chemistry are too high for CH$_4$ and too low for H$_2$O, when compared to retrieved values. That is, CH$_4$ needs to be quenched and H$_2$O needs to be enhanced to match with the retrieval result. 

This can be achieved by invoking vertical mixing at a quench pressure. Indeed, the retrieved abundances are consistent with the predicted values assuming a perfect vertical mixing for pressures lower than 50 bar (vertical solid lines in Fig. \ref{fig:chem_equi}). This indicates the case of chemical disequilibrium and a quench pressure that is higher than $\sim$10 bar.

\subsection{CO$_2$ \lowercase{vs.} no CO$_2$}
\label{sec:co2}

Similar to~\citet{Wang2020}, the peak of the retrieved CO$_2$ mass mixing ratio is $\sim$10$^{-4}$. While the posterior distribution has a long tail toward much lower abundance, the retrieved median of CO$_2$ mass mixing ratio is only two orders of magnitude lower than the retrieved median of CO mass mixing ratio. This is at odds with~\citet{Sorahana2012}, who predicted that CO$_2$ should be $\sim$4 orders of magnitude less abundant than CO for a brown dwarf with similar effective temperature although CO$_2$ may be produced through photochemistry~\citep{Moses2016}. We therefore perform another retrieval analysis where we fix the CO$_2$ mass mixing ratio to be 10$^{-10}$. The retrieved C and O abundances and C/O are consistent between the two models with varying and fixed CO$_2$ abundance. Neither model is preferred given the difference of $\ln$(Z) is only 1.3, where is Z is the Bayesian evidence. We therefore conclude that including/excluding CO$_2$ does not affect the retrieved C and O abundances and C/O.

\section{Summary}
\label{sec:summary}

\begin{itemize}
    \item Using photometric and spectroscopic data with resolving powers up to 35,000, we retrieve both elemental abundances and abundance ratios for HR 8799~c. We find that [C/H], [O/H], and C/O are 0.55$^{+0.36}_{-0.39}$, 0.47$^{+0.31}_{-0.32}$, and 0.67$^{+0.12}_{-0.15}$ at 68\% credible interval. Retrieval results are summarized in Table \ref{tab:mcmc_result} while a detailed discussion can be found in \S \ref{sec:discussion}. 
    \item The super-stellar C and O abundances and the stellar C/O ratio reveal potential formation pathways for HR 8799~c. Planet c, and likely the other planets in the system, {{formed within $\sim$1 Myr when there were sufficient solids to enhance the planetary atmosphere C and O abundances}}. The metal enrichment likely preceded or took place during the early phase of the inward migration to the planet current locations. More details can be found in \S \ref{sec:formation}. 
    \item The retrieval code is tested using a physically- and chemically-motivated PHOENIX synthetic spectrum with similar properties as those for HR 8799~c (\S \ref{sec:test}). We find that combining all available data, i.e., photometric data, and low- and high-resolution spectroscopic data, result in the most accurate inference of C and O abundances. Further discussions on retrieval uncertainties and systematics can be found in \S \ref{sec:low_high_comp}, \S \ref{sec:unc}, and \S \ref{sec:fixingLoggRadius}. 

\end{itemize}

In the future, we will extend this work to other planets in HR 8799 and test the consistency of formation pathway for the planets in the same system. We will take a similar approach to analyze other directly-imaged exoplanets with high-resolution spectroscopic data (e.g., KPIC) and lower resolution data that cover larger range of wavelength (e.g., the Jame Webb Space Telescope). 

\noindent
{\bf{Acknowledgments}} This work is supported by the National Science Foundation under Grant No. 2143400. {{We thank Andrew Youdin, Joan Najita, and Kaitlin Kratter for insightful discussions on gravitational instability vs. pebble accretion. }} We thank the anonymous referee for constructive comments and suggestions that significantly improve the paper.

Funding for KPIC has been provided by the California Institute of Technology, the Jet
Propulsion Laboratory, the Heising-Simons Foundation (grants \#2015-129, \#2017-318 and
\#2019-1312), the Simons Foundation (through the Caltech Center for Comparative Planetary
Evolution), and NSF under grant AST-1611623.

The data presented herein were obtained at the W. M. Keck Observatory, which is operated as a scientific partnership among the California Institute of Technology, the University of California and the National Aeronautics and Space Administration. The Observatory was made possible by the generous financial support of the W. M. Keck Foundation. The authors wish to recognize and acknowledge the very significant cultural role and reverence that the summit of Mauna Kea has always had within the indigenous Hawaiian community.  We are most fortunate to have the opportunity to conduct observations from this mountain.


We thank the Heising-Simons Foundation for supporting the workshop on combining high-resolution spectroscopy and high-contrast imaging for exoplanet characterization, where the idea originated on combining photometric data and spectral data of different resolutions. 

%






\bibliography{sample63}{}
\bibliographystyle{aasjournal}



\begin{deluxetable}{clllll}
\tablewidth{0pt}
\tablecaption{Summary of data sets \label{tab:datasets}}
\tablehead{
\colhead{\textbf{\#}} &
\colhead{\textbf{Telescope/}} &
\colhead{\textbf{Wavelength}} &
\colhead{\textbf{Resolution}} &
\colhead{\textbf{Mean}} &
\colhead{\textbf{Reference}} \\
\colhead{\textbf{}} &
\colhead{\textbf{Instrument}} &
\colhead{\textbf{Range ($\mu$m)}} &
\colhead{\textbf{}} &
\colhead{\textbf{SNR}} &
\colhead{\textbf{}}
}


\startdata
1 &Subaru/CHARIS & 1.16 - 2.37 & $\sim$20 & 13.8 & ~\citet{Wang2022b} \\
2 &Gemini/GPI & 1.50 - 2.37 & 45 - 80 & 15.3 & ~\citet{Greenbaum2018} \\
3 &Keck/OSIRIS & 1.96 - 2.38 & 5,000 & 42.2 & ~\citet{Konopacky2013} \\
4$^\ast$ &$L$-$M$ photometry& 3.04 - 4.67 & \nodata & 7.2 & ~\citet{Bonnefoy2016} \\
5 &Keck/KPIC & 2.291 - 2.482 & $\sim$35,000 & 0.6$^{\ast\ast}$ & ~\citet{Wang2021} \\
\hline
\multicolumn{6}{l}{HR 8799 c} \\
\hline
\enddata

\tablecomments{{$\ast$: A collection of data points from Keck/NIRC2, VLT/NaCo, Subaru/IRCS, and LBTI/LMIRCam~\citep{Galicher2011, Skemer2012, Skemer2014, Currie2014}. $\ast\ast$: CCF SNR$\sim$10~\citep{Wang2021}.}}

\end{deluxetable}

\begin{deluxetable}{clllll}
\tablewidth{0pt}
\tablecaption{C and O abundances for HR~8799 and planet c \label{tab:CO_abundances}}
\tablehead{
\colhead{\textbf{}} &
\colhead{\textbf{$\log{\epsilon_{\rm C}}$}} &
\colhead{\textbf{[C/H]}} &
\colhead{\textbf{$\log{\epsilon_{\rm O}}$}} &
\colhead{\textbf{[O/H]}} &
\colhead{\textbf{C/O}} 
}


\startdata
\multicolumn{6}{l}{HR 8799} \\
\hline
\citet{Wang2020} &8.54$\pm$0.12 & 0.04$\pm$0.12 & 8.81$\pm$0.14 & 0.08$\pm$0.14 & 0.54$^{+0.12}_{-0.09}$ \\
Solar$^\ast$ &8.43 & 0.00 & 8.69 & 0.00 & 0.55 \\
\hline
\multicolumn{6}{l}{HR 8799 c} \\
\hline
This work &8.98$^{+0.36}_{-0.39}$ & 0.55$^{+0.36}_{-0.39}$ & 9.16$^{+0.31}_{-0.32}$ & 0.47$^{+0.31}_{-0.32}$ & 0.67$^{+0.12}_{-0.15}$ \\
\enddata

\tablecomments{{$\ast$: We use solar elemental abundances from~\citet{Asplund2009}. We note that solar C and O abundances have been revised upward significantly in a recent work~\citep{Magg2022}. The revision affects values of [C/H] and [O/H] and the solar C/O, but will not affect results in this paper since we compare planet atmospheric abundances and abundance ratios to those of the host star. This highlights the importance of using stellar abundances instead of solar abundances when interpreting retrieval results. }}

\end{deluxetable}

\begin{deluxetable}{lccccccccc}
\rotate

\tabletypesize{\scriptsize}
\tablewidth{0pt}
\tablecaption{Summary of Retrieval Results.\label{tab:mcmc_result}}
\tablehead{
\colhead{\textbf{Parameter}} &
\colhead{\textbf{Unit}} &
\multicolumn{4}{c}{\textbf{PHOENIX}} &
\multicolumn{4}{c}{\textbf{HR 8799 c}} \\
\colhead{\textbf{}} &
\colhead{\textbf{}} &
\colhead{{SDS I}} &
\colhead{{SDS II}} &
\colhead{{SDS III}} &
\colhead{{Fixed $\log g$}} &
\colhead{{DS I}} &
\colhead{{DS II}} &
\colhead{\textbf{DS III}} &
\colhead{{Fixed CO$_2$}} 
}


\startdata
$\log$(g)                                & cgs                  & $      4.33^{     +0.13}_{     -0.14}$ &     $      4.53^{     +0.31}_{     -0.33}$ &     $      4.20^{     +0.21}_{     -0.25}$ &     $      3.50^{     +0.00}_{     -0.00}$ &     $      4.33^{     +0.39}_{     -0.42}$ &     $      4.32^{     +0.44}_{     -0.60}$ &     {\boldsymbol{$      4.17^{     +0.41}_{     -0.47}$}} &     $      4.30^{     +0.33}_{     -0.36}$\\
R$_P$                                    & R$_{\rm{Jupiter}}$   & $      1.30^{     +0.06}_{     -0.05}$ &     $      1.26^{     +0.07}_{     -0.06}$ &     $      1.36^{     +0.07}_{     -0.07}$ &     $      1.20^{     +0.00}_{     -0.00}$ &     $      1.06^{     +0.06}_{     -0.07}$ &     $      0.95^{     +0.05}_{     -0.04}$ &     {\boldsymbol{$      1.01^{     +0.09}_{     -0.08}$}} &     $      0.96^{     +0.06}_{     -0.05}$\\
$\Delta_y$                               & \nodata              & $     -0.01^{     +0.31}_{     -0.31}$ &     $     -0.04^{     +0.06}_{     -0.06}$ &     $     -0.08^{     +0.10}_{     -0.09}$ &     $     -0.09^{     +0.09}_{     -0.08}$ &     $      0.01^{     +0.30}_{     -0.31}$ &     $      0.24^{     +0.03}_{     -0.03}$ &     {\boldsymbol{$      0.14^{     +0.06}_{     -0.06}$}} &     $      0.16^{     +0.05}_{     -0.05}$\\
$\log$(mr$_{\rm{H}_2\rm{O}}$)            & \nodata              & $     -2.24^{     +0.05}_{     -0.05}$ &     $     -2.28^{     +0.13}_{     -0.11}$ &     $     -2.27^{     +0.08}_{     -0.07}$ &     $     -2.50^{     +0.04}_{     -0.03}$ &     $     -2.18^{     +0.21}_{     -0.22}$ &     $     -2.07^{     +0.19}_{     -0.25}$ &     {\boldsymbol{$     -2.22^{     +0.24}_{     -0.25}$}} &     $     -2.14^{     +0.20}_{     -0.20}$\\
$\log$(mr$_{\rm{C}\rm{O}}$)              & \nodata              & $     -1.95^{     +0.12}_{     -0.12}$ &     $     -2.14^{     +0.33}_{     -0.33}$ &     $     -2.05^{     +0.20}_{     -0.21}$ &     $     -2.17^{     +0.18}_{     -0.22}$ &     $     -1.75^{     +0.34}_{     -0.37}$ &     $     -1.30^{     +0.19}_{     -0.28}$ &     {\boldsymbol{$     -1.72^{     +0.37}_{     -0.39}$}} &     $     -1.60^{     +0.31}_{     -0.33}$\\
$\log$(mr$_{\rm{C}\rm{O}_2}$)            & \nodata              & $     -3.79^{     +0.51}_{     -3.49}$ &     $     -3.51^{     +0.75}_{     -3.34}$ &     $     -5.45^{     +1.84}_{     -2.85}$ &     $     -7.24^{     +1.83}_{     -1.71}$ &     $     -2.85^{     +0.34}_{     -0.50}$ &     $     -6.20^{     +1.95}_{     -2.35}$ &     {\boldsymbol{$     -3.50^{     +0.67}_{     -3.44}$}} &     $    -10.00^{     +0.00}_{     -0.00}$\\
$\log$(mr$_{\rm{C}\rm{H}_4}$)            & \nodata              & $     -4.24^{     +0.13}_{     -0.12}$ &     $     -4.23^{     +0.27}_{     -0.24}$ &     $     -4.31^{     +0.17}_{     -0.17}$ &     $     -4.67^{     +0.09}_{     -0.10}$ &     $     -4.20^{     +0.33}_{     -0.36}$ &     $     -6.66^{     +1.65}_{     -2.07}$ &     {\boldsymbol{$     -4.39^{     +0.38}_{     -0.44}$}} &     $     -4.34^{     +0.33}_{     -0.43}$\\
t$_{\rm{int}}$                           & K                    & $      1850^{       +67}_{       -66}$ &     $      1642^{       +95}_{      -113}$ &     $      1809^{      +101}_{       -92}$ &     $      1983^{      +150}_{       -86}$ &     $      1380^{       +99}_{       -79}$ &     $      1494^{      +180}_{       -92}$ &     {\boldsymbol{$      1421^{       +92}_{       -72}$}} &     $      1443^{      +100}_{       -79}$\\
$\Delta\lambda_{\rm{KPIC}}$              & $\AA$                & $     -0.54^{    +62.52}_{    -62.42}$ &     $     -0.13^{     +0.33}_{     -0.33}$ &     $      0.49^{    +54.02}_{    -27.10}$ &     $      0.16^{    +37.63}_{     -5.30}$ &     $     -1.32^{    +61.58}_{    -61.40}$ &     $      3.04^{     +0.11}_{     -0.12}$ &     {\boldsymbol{$      3.04^{     +0.52}_{     -0.54}$}} &     $      3.03^{     +0.27}_{     -0.28}$\\
$\Delta\lambda_{\rm{OSIRIS}}$            & $\AA$                & $     -0.48^{     +0.25}_{     -0.37}$ &     $      1.00^{    +62.41}_{    -63.05}$ &     $     -0.61^{     +0.44}_{     -0.50}$ &     $     -0.81^{     +0.30}_{     -0.42}$ &     $     -4.85^{     +0.70}_{     -0.76}$ &     $     -1.63^{    +63.16}_{    -60.56}$ &     {\boldsymbol{$     -4.80^{     +0.86}_{     -1.17}$}} &     $     -4.58^{     +0.57}_{     -0.57}$\\
$\Delta T$ between 100 and 32 bar        & K                    & $       736^{      +780}_{      -509}$ &     $       644^{      +783}_{      -462}$ &     $      1019^{      +852}_{      -666}$ &     $      1242^{      +767}_{      -756}$ &     $      1102^{      +822}_{      -719}$ &     $      1193^{      +792}_{      -748}$ &     {\boldsymbol{$      1126^{      +811}_{      -728}$}} &     $      1163^{      +806}_{      -744}$\\
$\Delta T$ between 32 and 10 bar         & K                    & $       126^{      +172}_{       -92}$ &     $       150^{      +212}_{      -109}$ &     $       265^{      +368}_{      -193}$ &     $       457^{      +499}_{      -316}$ &     $       685^{      +799}_{      -538}$ &     $       740^{      +680}_{      -506}$ &     {\boldsymbol{$       557^{      +721}_{      -402}$}} &     $       728^{      +753}_{      -530}$\\
$\Delta T$ between 10 and 3.2 bar        & K                    & $        62^{       +75}_{       -44}$ &     $       105^{      +125}_{       -75}$ &     $       112^{      +132}_{       -80}$ &     $        93^{      +106}_{       -68}$ &     $       360^{      +653}_{      -297}$ &     $       289^{      +432}_{      -207}$ &     {\boldsymbol{$       216^{      +513}_{      -156}$}} &     $       383^{      +564}_{      -295}$\\
$\Delta T$ between 3.2 and 1 bar         & K                    & $       806^{       +99}_{      -116}$ &     $       497^{      +156}_{      -146}$ &     $       659^{      +146}_{      -165}$ &     $       280^{      +148}_{       -98}$ &     $       112^{      +127}_{       -72}$ &     $       111^{      +158}_{       -76}$ &     {\boldsymbol{$       104^{      +115}_{       -68}$}} &     $       128^{      +137}_{       -81}$\\
$\Delta T$ between 1 and 0.1 bar         & K                    & $       248^{      +160}_{      -137}$ &     $       565^{      +271}_{      -290}$ &     $       413^{      +264}_{      -213}$ &     $       969^{       +22}_{       -36}$ &     $       115^{      +102}_{       -75}$ &     $       338^{       +86}_{       -90}$ &     {\boldsymbol{$       152^{       +79}_{       -82}$}} &     $       119^{       +86}_{       -73}$\\
$\Delta T$ between 0.1 bar and 1 mbar    & K                    & $       380^{      +343}_{      -258}$ &     $       477^{      +321}_{      -306}$ &     $       413^{      +351}_{      -275}$ &     $       247^{      +252}_{      -164}$ &     $       659^{      +226}_{      -288}$ &     $       640^{      +231}_{      -289}$ &     {\boldsymbol{$       592^{      +257}_{      -305}$}} &     $       578^{      +275}_{      -320}$\\
$\Delta T$ between 1 mbar and 10 nbar    & K                    & $       507^{      +306}_{      -307}$ &     $       496^{      +318}_{      -310}$ &     $       495^{      +315}_{      -316}$ &     $       509^{      +304}_{      -309}$ &     $       496^{      +308}_{      -310}$ &     $       497^{      +312}_{      -307}$ &     {\boldsymbol{$       496^{      +316}_{      -301}$}} &     $       496^{      +315}_{      -309}$\\
f$_{\rm{blur}}$                          & km$\cdot$s$^{-1}$    & $     50.45^{    +31.35}_{    -29.89}$ &     $     22.58^{    +13.18}_{     -9.51}$ &     $     67.53^{    +21.82}_{    -27.48}$ &     $     68.51^{    +20.77}_{    -26.34}$ &     $     50.45^{    +30.50}_{    -30.58}$ &     $     12.08^{     +4.23}_{     -2.89}$ &     {\boldsymbol{$     20.35^{    +28.09}_{    -10.59}$}} &     $     14.01^{    +11.13}_{     -6.13}$\\
$\log$(mr$_{\rm{MgSiO}_3}$)              & \nodata              & $     -4.47^{     +1.24}_{     -0.42}$ &     $     -4.85^{     +1.17}_{     -0.98}$ &     $     -3.69^{     +1.02}_{     -1.02}$ &     $     -3.11^{     +0.37}_{     -0.40}$ &     $     -3.89^{     +1.06}_{     -3.57}$ &     $     -3.68^{     +0.97}_{     -1.34}$ &     {\boldsymbol{$     -4.76^{     +1.67}_{     -3.06}$}} &     $     -3.80^{     +1.07}_{     -2.76}$\\
$\log$(K$_{zz}$)                         & cm$2\cdot$s$^{-1}$   & $      5.75^{     +2.60}_{     -0.57}$ &     $      7.30^{     +1.78}_{     -1.67}$ &     $      7.33^{     +1.85}_{     -1.82}$ &     $      5.27^{     +0.40}_{     -0.20}$ &     $      7.81^{     +1.25}_{     -1.33}$ &     $      7.73^{     +1.37}_{     -1.46}$ &     {\boldsymbol{$      7.78^{     +1.38}_{     -1.50}$}} &     $      7.79^{     +1.34}_{     -1.42}$\\
$f_{sed}$                                & \nodata              & $      4.65^{     +0.25}_{     -0.41}$ &     $      4.16^{     +0.59}_{     -0.88}$ &     $      4.47^{     +0.38}_{     -0.59}$ &     $      4.55^{     +0.31}_{     -0.51}$ &     $      2.41^{     +1.19}_{     -1.05}$ &     $      2.69^{     +1.26}_{     -1.16}$ &     {\boldsymbol{$      2.60^{     +1.33}_{     -1.29}$}} &     $      2.65^{     +1.25}_{     -1.18}$\\
$\sigma_g$                               & \nodata              & $      2.40^{     +0.44}_{     -0.65}$ &     $      2.10^{     +0.60}_{     -0.65}$ &     $      2.18^{     +0.58}_{     -0.66}$ &     $      2.44^{     +0.39}_{     -0.57}$ &     $      2.03^{     +0.60}_{     -0.59}$ &     $      2.03^{     +0.61}_{     -0.60}$ &     {\boldsymbol{$      2.03^{     +0.62}_{     -0.59}$}} &     $      2.02^{     +0.63}_{     -0.60}$\\
$[$C/H$]$                                & dex                  & $      0.31^{     +0.12}_{     -0.12}$ &     $      0.14^{     +0.33}_{     -0.32}$ &     $      0.21^{     +0.20}_{     -0.20}$ &     $      0.09^{     +0.18}_{     -0.22}$ &     $      0.54^{     +0.34}_{     -0.36}$ &     $      0.95^{     +0.19}_{     -0.28}$ &     {\boldsymbol{$      0.55^{     +0.36}_{     -0.39}$}} &     $      0.65^{     +0.31}_{     -0.33}$\\
$[$O/H$]$                                & dex                  & $      0.31^{     +0.08}_{     -0.08}$ &     $      0.20^{     +0.23}_{     -0.18}$ &     $      0.24^{     +0.13}_{     -0.12}$ &     $      0.06^{     +0.12}_{     -0.12}$ &     $      0.49^{     +0.28}_{     -0.29}$ &     $      0.80^{     +0.18}_{     -0.27}$ &     {\boldsymbol{$      0.47^{     +0.31}_{     -0.32}$}} &     $      0.56^{     +0.27}_{     -0.27}$\\
C/O                                      & \nodata              & $      0.56^{     +0.06}_{     -0.06}$ &     $      0.48^{     +0.14}_{     -0.15}$ &     $      0.52^{     +0.10}_{     -0.10}$ &     $      0.58^{     +0.09}_{     -0.12}$ &     $      0.63^{     +0.11}_{     -0.13}$ &     $      0.78^{     +0.06}_{     -0.07}$ &     {\boldsymbol{$      0.67^{     +0.12}_{     -0.15}$}} &     $      0.69^{     +0.10}_{     -0.12}$\\
\enddata

\tablecomments{{a: We report the median of posterior distribution and error bars correspond to the difference of the median and the 68\% credible interval. b: We adopt solar elemental abundances from~\citet{Asplund2009}}}

\end{deluxetable}

\end{CJK*}
 
\end{document}